\documentstyle[epsf,epsfig]{l-aa}

\begin{document}

\thesaurus{04         
              (11.09.3; 
	       09.03.1; 
               11.12.1;  
               12.04.1)} 

\title{High-velocity clouds as dark matter in the Local Group}


\author{M. L\'opez--Corredoira, J. E. Beckman, E. Casuso}

\offprints{martinlc@iac.es}

\institute{Instituto de Astrof\'\i sica de Canarias,
              E-38200 La Laguna, Tenerife, Spain}

\date{Received 29 June 1999; accepted 16 August 1999}

\maketitle 

\markboth{L\'opez-Corredoira et al.}{HVCs as dark matter}

\begin{abstract}
The High-Velocity Clouds (HVCs) observed in the Galactic
neighbourhood, have been proposed
to be remnants of the formation of the galaxies in the Local Group, having
distances, and thus masses, predominantly of dark matter, 
considerably larger than hitherto hypothesized.
This hypothesis is plausibly supported by observational evidence that
their kinematical centre is the Local Group barycentre. Evolutionary models
to account for the evolution of the light elements in the Galaxy demand 
infall of metal poor gas to the plane, which could well be supplied by these
HVCs. Modelling the time dependence of this infall, taking into account that an accreting galaxy 
shows an increasing cross-section to the infalling clouds, and produces 
increasing mean infall velocities, we deduce that the HVCs must currently
represent at least around one half of the total mass of the Local Group, given
that the accretion rate, as inferred from chemical evolution, has not decreased
significantly during the disc lifetime. This fraction is consistent with 
dynamical estimates of the relative masses of the Local Group as a whole
and its constituent galaxies. The HVCs may thus form a significant constituent of 
baryonic, and of non-baryonic, dark matter. 

\keywords{Intergalactic medium -- ISM: clouds -- Local Group -- dark matter}
\end{abstract}

\section {Introduction}

High-velocity clouds (HVCs) consist of neutral hydrogen at radial velocities
approximately $|v_{\rm LSR}|>90$ km/s,
incompatible with a simple model of differential galactic rotation
(Wakker \& van Woerden 1997). Their range of radial
LSR velocities extends indeed up to -500 km/s$<v_{\rm LSR}<$+300 km/s.
Several hypotheses try to explain the origins
of the HVCs (see review in Wakker \& van Woerden 1997;
Wakker et al. 1999b). At least three different sources appear to be needed: 
one for the Magellanic Stream
and related clouds, one for the Outer Arm Extension, and at
least one for the ``other HVCs'' (Wakker \& van Woerden 1997).

One hypothesis, recently put forward by Blitz et al. (1999, hereafter BL99)
as the most plausible one to explain ``other'' HVCs, claims that they
are remnants of the formation of the Local Group at a scale distance of 1 Mpc,
i.e. far from the Galaxy. These clouds are falling towards the 
Local Group barycenter and some of them will be accreted by the Milky Way
if they move close enough to it.

Here, we hypothesize that these clouds may constitute
a major fraction of the mass of the Local Group.
The fraction of sky covered by HVCs with $|v_{\rm LSR}|>100$ km/s and
$N({\rm HI})>2\times 10^{18}$ cm$^{-2}$ is 8\%, excluding the Outer Arm Extension
and the Magellanic Stream (Wakker 1991). If they are relatively
distant, as the estimated mass of a cloud is 
proportional to $d^2$ ($d$ is the distance), they 
could represent a larger contribution to the mass of the Local Group than 
hitherto assumed.

\section{Accretion theory}

The view of HVCs as gas being accreted by the Galaxy was first suggested
by Oort (1966, 1970), who postulated that gas left over from the formation
of the Galaxy is now reaching the Galactic disc. During their 
approach, the clouds are heated and recooled forming the HVCs at
heights of 1 to 3 kpc. A simple model with the Galaxy accreting
these clouds encounters directly many problems; for instance, it cannot justify
observations with positive velocities. 
The possibility of an extragalactic origin of these clouds
associated with the Local Group
was discussed by Verschuur (1969, 1975),
Einasto et al. (1976), Eichler (1976),
Giovanelli (1980, 1981), Arp (1985), Bajaja et al. (1987) and others, 
but their models apparently failed to fully explain the observations.

The difficulties of this hypothesis were solved in BL99 aided by new
observational evidence. The kinematic anomalies 
are explained in terms of the infall of remnants 
of the formation of the Local Group towards its barycentre. 
A typical cloud at a distance of 1 Mpc has a diameter of 30 kpc
and contains $3.5\times 10^7$ M$_\odot$ of HI gas, 
and a total mass of $\sim 3\times 10^8$ M$_\odot$ (given 85\% of dark mass). 
Its density would be $<n_{\rm HI}>\approx 10^{-4}$ cm$^{-3}$.
Braun \& Burton (1999) analyzed a sample of HVCs similar to that
of BL99, but chose isolated clouds, avoiding the contamination
of some complexes related to the Magellanic Stream
or the Outer Arm Extension, finding a size of 15 kpc and a similar
HI mass. The dynamics of the Local Group is dominated by the attraction of
M31 (with twice the mass of the Milky Way)
and the Milky Way. When a cloud is within 100 kpc comoving distance
of one of these galaxies it can be taken as accreted.
The simulated spatial and kinematic distributions resemble 
the observed distributions rather well. Most of the
HVCs are located either near the direction of M31, towards the
barycenter of the Local Group, or in the antibarycenter direction.
The distribution of velocities with angular distance
from the solar apex is that expected for Local Group
barycentre infall (Braun \& Burton 1999).

Low metallicity is observed in some HVCs. Sembach et al. (1999)
observed some highly ionized HVCs whose ionic ratios are well explained 
via photoionization by extragalactic background radiation combined with
some ultraviolet starlight. The observations by
Tufte et al. (1998) of a set of HVCs also suggest 
photoionization. Hence, these clouds must have low density
($n_{\rm H} \sim 10^{-4}$ cm$^{-3}$), and be bigger than a few kpc
and mainly ionized ($n_{\rm HI}/n_{\rm H} \sim 10^{-3}$), which indicate that they
must be far away from the neutral Galactic gas. These
clouds would have total masses of $\sim 10^8$ M$_\odot$. 
If the clouds are intergalactic
in nature, their metallicities could well be $[Z/H]\sim -1$ or lower.
Low metallicities were also detected in an HVC through sulphur abundance 
measurements, yielding
$[S/H]=0.094$ times solar, much lower than the solar value (Wakker et al. 1999a).
This metallicity supports the extragalactic
origin at least for this object because gas ejected from the current 
Galaxy would share its near-solar metallicity.

As an additional datum, it is observed that the
HVCs properties are similar to those found 
for high redshift ``Ly$\alpha $ forest''
clouds (BL99). This implies that this type of clouds
is also abundant at very large distances from the Local Group and that 
they are present in the primordial scenarios of galaxy formation.

\section{Chemical evolution of the Galaxy}

If HVCs are extragalactic, the chemical evolution of the
Galactic disc is strongly affected by continual but episodic infall of metal-poor gas from 
HVCs which mixes slowly with the rest of the interstellar medium.

The infall rate of low metallicity gas should
be increasing now, or at least remain constant on timescales compared to
the Hubble time. The reasons which sustain 
this hypothesis are based on a interpretation of the chemical evolution
in the local Galactic disc:

\begin{itemize}

\item The G-dwarf distribution in the solar neighbourhood 
(Rocha-Pinto \& Maciel 1996) with a sharp rise
in numbers close to $[Fe/H]=-0.2$ requires either a sharp increase in
the SFR (stellar formation rate) at that metallicity, 
or the accumulation of stars formed at
different epochs but at the same metallicity and both can be explained using
a rising or near constant infall rate of metal-poor gas.

\item Star formation in the past 5 Gyr has a long-term average slow-rise
according to stellar activity data
(Vaughan \& Preston 1980, Barry 1988).

\item The $^9$Be abundance plotted versus iron abundance has a loop-back close to
$[Fe/H]=-0.2$, implying a fall in Fe abundance while $^9$Be abundance increases,
which can be well explained with a rising infall rate (Casuso \& Beckman 1997).

\end{itemize}

An inference presented in BL99 is that the accretion of
gas in the Milky Way is decreasing. The present value would be
7.5 M$_\odot$/yr. However, in their simple dynamical model, 
they do not take into account the increasing gravitational
potential of the discs due to the accretion itself.
The increasing mass of our Galaxy could
counterbalance the declining density of the baryonic material in the
intergalactic medium leading to an increasing accretion rate of material into
the Galaxy. This will be discussed quantitatively in the next section.

\section{Total mass of the clouds in the Local Group}

The accretion rate depends on the spatial 
density of the clouds ($\langle\rho _{\rm clouds}\rangle$), 
their mean velocities ($\langle v\rangle$) 
and the cross section of galaxies to trap a cloud
($\Sigma _{\rm g}$). These three factors govern the dynamics:

\begin{equation}
\frac{dM_{\rm g}}{dt}\propto \langle \rho _{\rm clouds}\rangle 
\langle v \rangle 
\Sigma_{\rm g}
,\end{equation}
where $M_{\rm g}$ is the mass of the accreting galaxies plus their
neighbour satellites; the main components are the Milky Way and M31.

BL99 take the second and third factors as constants but
this is too simplistic. The three factors are respectively:

\begin{description}

\item[Density.] The density can be expressed simply via:

\begin{equation}
\langle\rho _{\rm clouds}\rangle=\frac{M-M_{\rm g}}{V}
,\end{equation}
where $M$ is the total mass of the Local group (constant, assuming that
the local group does not exchange mass with any external reservoir) 
which includes the mass of the clouds and the galaxies ($M_{\rm g}$). 
Any ionized intercloud medium would make an extra contribution
to the mass of the Local Group; although we do not include it explicitly here, 
its gravitational behaviour would be akin to that of the clouds.
$V$ is the physical volume of the Local Group, which is constant 
(we can assume that the Hubble
expansion is neutralized on the scale of the Local Group 
since the cluster is gravitationally bound and collapse, if any, is very
slow).

\item[Velocity.]
The velocity depends on the masses of the accreting galaxies.
When the cloud approaches a centre of accretion, 
the average velocity of the cloud with mass $m_{\rm c}$ is:

\begin{equation}
\frac{1}{2}m_{\rm c}v_1(r_1)^2 = s\frac{GM_1m_{\rm c}}{r_1}
\label{m1}
,\end{equation}
where $r_1$ is the distance to the accreting galaxy of mass $M_1$ 
and $v_1=\left|\frac{d\vec{r_1}}{dt}\right|$, the velocity with respect to this object. 
$<s>$ depends on the mean energy of the clouds; $s=1$ if
they have escape velocity, $s=\frac{1}{2}$ for a virialized system; etc.
Masses of other objects have a negligible effect when the cloud is quite 
close to a galaxy, so, roughly, 
the velocity is proportional to $M_1^{1/2}$.
In our case, there are two accreting centres and the relation is
more complicated. Eq. (\ref{m1}) applies near
the mass $M_1$. Near the mass $M_2$, the average velocity 
of the clouds is:

\begin{equation}
\frac{1}{2}m_{\rm c}v_2(r_2)^2 = s\frac{GM_2m_{\rm c}}{r_2}
\label{m2}
,\end{equation}
where $r_2$ is the distance to the accreting galaxy of mass $M_2$ 
and $v_2=\left|\frac{d\vec{r_2}}{dt}\right|$. Therefore,
the velocity is proportional to $M_2^{1/2}$.

In the region dominated by the barycentre rather
than each individual galaxy, we can take 
$M_{\rm g}+C(M-M_{\rm g})$ as the equivalent mass of the centre of attraction.
The first term is due to the galaxies ($M_{\rm g}=M_1+M_2$) and the second
is the mass of the clouds which are inside the spherical region centred
on the barycentre, and radius the distance to it.
Since the distance is a variable, $C$ (the fraction of the mass
in clouds in that region: between 0 and 1) is variable,
but we can take an average value for $C$ to tipify infall of the clouds.  
The continuous increment of kinetic energy during infall is the
sum of terms due to the galaxies and the clouds; hence, the total 
increment is the sum of two terms: one proportional to $M_{\rm g}$ and another
one proportional to $(M-M_{\rm g})$.

Near to the galaxies, making the rasonable assumption that $M_1/M_{\rm g}$ 
and $M_2/M_{\rm g}$ are 
constants, i.e. the ratio of the masses of the two galaxies does not change 
and the velocity of the galaxies with respect the barycentre is negligible 
compared with that of the clouds, the mean infall 
velocity ($v_1=v_2=v_{\rm bar}$) will be proportional to $M_{\rm g}^{1/2}$, i.e. $C=0$.
The mean velocity is, in general, proportional to $\sim (M_{\rm g}+C(M-M_{\rm g}))^{1/2}$
where $C$ is an averaged quantity along the path of the clouds and lies 
between 0 and 1
($v^2$ is proportional to $(C_1M_{\rm g}+C_2(M-M_{\rm g}))$, since in the proximity
of the galaxy $i$, $C_1=M_{\rm i}/M_{\rm g}$, while on the rest of the path $C_1=1$, 
but this is solved by dividing the expression for $v^2$
by the average $C_1$; so $C=C_2/C_1$).
 
The velocity may be slightly reduced by interaction 
with the disc, via collision of the high-velocity gas
with the stationary disc gas (see calculations in 
Tenorio-Tagle et al. 1987, 1988; Comeron \& Torra 1992, 1994). 
If drag forces dominate, the velocity becomes
proportional to the square root of the column density. Benjamin \& Danly (1997)
found that intermediate-velocity clouds with low column densities may well 
reach terminal velocity, but HVCs should be hardly slowed down.
As their time of interaction with the disc is relatively short,
the effects of drag are negligible and the velocity can be calculated without
taking them into account.

\item[Collision cross section of galaxies for the clouds.]
This is proportional to the square of the disc radius of the
galaxy. In an essentially two-dimensional disc, such as that of a 
spiral galaxy, we can assume a dependence of the radius
on the mass of form $L\propto
M^{1/2}$. This is in fact found observationally. In the plot of the radii
vs. the masses of a sample of S$_{\rm b}$ galaxies (Campos-Aguilar et al. 1993),
a relation $L\propto M^{0.49\pm 0.06}$ is fitted. Since M31 is an 
S$_{\rm b}$ galaxy and the Milky Way is an S$_{\rm bc}$, we can use this relationship
for these objects and take the cross section to be proportional to $L^2$, i.e.
proportional to $M_{\rm g}$.

\end{description}

Therefore, the accretion rate follows:

\begin{equation}
\frac{dM_{\rm g}}{dt}=\frac{K}{V}(M-M_{\rm g})[M_{\rm g}+C(M-M_{\rm g})]^{1/2}
M_{\rm g}\label{infall1}
,\end{equation}
where $K$ is a positive constant.

The factor $(M-M_{\rm g})>0$ falls with time thereby reducing
the accretion rate for the Milky Way, but $(M_{\rm g}+C(M-M_{\rm g})^{1/2}
M_{\rm g}$ 
increases yielding an increase with time.
Whether the net rate increases will depend on the ratio of the masses
of the major galaxies to the integrated mass available for accretion in the
clouds. 

If the accretion rate is increasing or at least constant, 
as argued in the previous section,
the minimum limit implied for the fractional mass of the HVCs in the Local 
Group can be evaluated
and compared with observational parameters.
The condition for the rate to be increasing with the time at the present
epoch is that its derivative
be greater than zero:

\[
\frac{d^2M_{\rm g}}{dt^2}=\frac{K}{V}(-[M_{\rm g}+C(M-M_{\rm g})]^{1/2}M_{\rm g}
\]\[+
(M-M_{\rm g})M_{\rm g}(1-C)[M_{\rm g}+C(M-M_{\rm g})]^{-1/2}/2
\]\[ 
+(M-M_{\rm g})[M_{\rm g}+C(M-M_{\rm g})]^{1/2})
\frac{dM_{\rm g}}{dt}>0
\]\begin{equation}
\Rightarrow 
M_{\rm g}<\frac{4C}{\sqrt{9C^2-2C+9}-3+7C}M
,\end{equation}
\begin{equation}
M_{\rm clouds}=M-M_{\rm g}>\frac{\sqrt{9C^2-2C+9}-3+3C}{\sqrt{9C^2-2C+9}-3+7C}M
.\end{equation}
This gives us a minimum fraction of mass for the clouds within the local Group
which is between 40\% and 50\%, depending on the value of $C$ (between 0 and 1).
Although this result is approximate, it is qualitatively
plausible: a low fraction of mass in clouds would not
yield increasing accretion as the mean cloud density would fall
more quickly than could be compensated by the increasing
gravitational attraction of the accreting galaxies.

This result is in very fair agreement with the dynamical 
estimates of the total mass of the local group,
$\sim 3\times 10^{12}$ M$_\odot$ (Byrd et al. 1994), 
in which well over the half
of mass is not in the baryonic masses of detectable galaxies. 
The HVCs would constitute much of the non-galactic mass of the Local Group, they would
be at least a significant part of the dark matter in the Local Group. 
One can speculate that the dark matter
(the difference between the total mass of the
cluster and the sum of the masses of the individual galaxies) 
in all clusters of galaxiesis due in significant degree to such clouds.

We can infer that the minimum mass of the clouds
in the Local Group is $\sim 1.2-1.5\times 10^{12}$ M$_\odot$, i.e. around
5000 clouds with an average mass of $\sim 3\times 10^{8}$ M$_\odot$.
Since the Milky Way mass, from its rotation 
curve within the inner 15 kpc, is around $2\times 10^{11}$ M$_\odot$ 
(Honma \& Sufue 1996), M31 has around twice this mass, and the rest of the
galaxies also contribute a little, we can postulate that  
$\sim 2\times 10^{12}$ M$_\odot$ is the maximum mass of the clouds
given the above value of the total mass for the Local Group.

The clouds are not homogeneouly distributed, and most of them should be at
a distance of $d\sim 1$ Mpc. The mean column density expected from them
would be:

\begin{equation}
<N({\rm HI})>_{\rm sky}=\frac{M_{\rm clouds}f\ N_{\rm A}}{4\pi d^2}
,\end{equation} 
where $f$ is the fraction of gaseous hydrogen ($f=0.15$ according to BL99) 
and $N_{\rm A}$ is Avogadro's number. 
The result is $<N({\rm HI})>_{\rm sky}\sim
2\times 10^{18}$ cm$^{-2}$, whose order of magnitude
is in agreement with observations (Wakker 1991).
The column density extends over the range $2\times 10^{17}$ cm$^{-2}
<N({\rm HI})<2\times 10^{20}$ cm$^{-2}$, and the observed average would be somewhat 
lower than $3\times 10^{18}$ cm$^{-2}$ for $|v_{\rm LSR}|>100$ km/s; 
although this difference might be due to the non-inclusion of intermediate velocity clouds as 
well to scatter in the parameters used ($M_{\rm clouds}$, $f$, $d$).

\section{Dark matter in HVCs and galaxies}

Most of the HVC mass is dark matter, around 85\% of the mass of the clouds, 
i.e. more than a half of the total mass
in the Local Group. The nature of this dark matter is unknown.
The dark matter in the HVCs could be some kind of baryonic matter; at least,
this is compatible with the nucleosynthesis values: the baryonic matter is 
limited to 0.018$<\Omega _{\rm b}h^2<$0.022 (Schramm \& Turner 1998) while the observed
local stellar density is a 17\% of that (Fukugita et al. 1998).

This does not mean that all dark matter has to be baryonic, but the total
mass of these clouds could be in baryons. Baryonic dark matter
exists (Silk 1996) and a fraction
of half of the mass of the Local Group is not excluded
by primordial nucleosynthesis of the light elements.
Very low mass stars, cold hydrogen and other exotic matter
are some candidates. Models derived in part from the observations
predict a scenario in which molecular clouds and dark clusters of
MACHOs constitute the halo (De Paolis et al. 1995), or very cold molecular
hydrogen in the disc (Pfenninger et al. 1994).
These may also constitute scenarios for the dark matter
in the HVCs.

Since a significant fraction of the present disc 
material proceeds from HVC accretion,
part of the dark matter in the Galactic halo and disc is the same
kind of material as the dark matter of the HVCs although the fraction
may be different.
From Eq. (\ref{infall1}), the accretion rate can be derived
for any time. For instance, assuming a present rate in the Milky Way
of 7.5 M$_\odot$/yr (BL99), 
twice this for M31 and the total mass of the Local Group
$\sim 3\times 10^{12}$ M$_\odot$, we find that less than 
30\% of $M_{\rm g}$ has come from the
accretion of HVCs during the last 15 Gyr. 
This is much less than the predictions
by BL99, who conclude that practicaly all the mass of the two major
disc galaxies should be due to the accretion of HVCs, mainly during
early epochs.
Unless its non-baryonic nature implies a process which favours only the
retention of the baryonic components, this would imply that less 
than 25\% of the accreted matter of the
Milky Way or M31 may be constituted by the kind of 
dark matter which is present in the HVCs. To this fraction, one should,
of course, add the dark matter which entered into the constitution of the
galaxies when they formed.

\section{Conclusions}

Using essentially the BL99 model for HVCs
and a non-decreasing accretion rate  
of these clouds with low metallicity onto the Milky Way, we
have deduced that their present abundance in the Local Group 
must be greater than some 40\% to 50\% of its total mass. 
This is in agreement with the dynamical
measurements of the total mass in the Local Group and may account
for a significant fraction of the baryonic dark matter, 
although a major fraction of non-baryonic matter 
within these clouds is not excluded.
The calculations shown here are a first approximation, in which collapse or expansion 
of the Local Group was not taken into account but, the volume variation, if any,
should be slow enough for us to neglect its effects.

At present, the nature of this dark matter (around 85\% of the total
mass of the clouds) cannot be determined, but
the hypothesis of an important contribution to the total mass of the
Local Group may be a clue to towards finding it. 
Rotation curve anomalies and warps (Jiang \& Binney 1999)
in spiral galaxies could also be due to the existence of these HVCs,
as a response to the torque the HVC system can exert on the disc of a large
spiral.

\acknowledgements
We thank the referee W. B. Burton for some helpful comments.


\begin{thebibliography}{99}

\bibitem{} Arp H., 1985, AJ 90, 1012

\bibitem{} Bajaja E., Morras R., P\"oppel W. G. L., 1987,
Pub. Astr. Inst. Czech. Ac. Sci. 69, 237

\bibitem{} Barry D. C., 1988, ApJ 334, 436

\bibitem{} Benjamin R. A., Danly L., 1997, ApJ
481, 764

\bibitem{} Blitz L., Spergel D., Teuben P.,
Hartmann D., Burton W. B., 1999, ApJ 514, 818 (BL99)

\bibitem{} Braun R., Burton W. B., 1999, A\&A 341, 437

\bibitem{} Byrd G., Valtonen M., McCall M.,
Innamen K., 1994, AJ 107, 2055

\bibitem{} Campos-Aguilar A., Prieto M.,
Garc\'\i a C., 1993, A\&A 276, 16

\bibitem{} Casuso E., Beckman J. E., 1997,
ApJ 475, 155

\bibitem{} Comeron F., Torra J., 1992,
A\&A 261, 94

\bibitem{} Comeron F., Torra J., 1994,
A\&A 281, 35

\bibitem{} De Paolis F., Ingrosso G.,
Jetzer P., Roncadelli M., 1995, A\&A 295, 567

\bibitem{} Eichler D., 1976, ApJ 208, 694

\bibitem{} Einasto J., Haud U., Foeveer M., Kaasik A., 1976, MNRAS 177, 357

\bibitem{} Fukugita M., Hogan C. J., Peebles P. J. E., 
1998, ApJ 503, 518

\bibitem{} Giovanelli R., 1980, AJ 85, 1155

\bibitem{} Giovanelli R., 1981, AJ 86, 1468

\bibitem{} Honma M., Sofue Y., 1996,
PASJ 48, L103

\bibitem{} Jiang I. G., Binney J., 1999,
MNRAS 303, L7

\bibitem{} Oort J. H., 1966, Bull. Astron.
Inst. Neth. 18, 421

\bibitem{} Oort J. H., 1970, A\&A 7, 381

\bibitem{} Pfenninger D., Combes F.,
Martinet L., 1994, A\&A 285, 79

\bibitem{} Rocha-Pinto H. J., Maciel W. J, 1996,
MNRAS 279, 447

\bibitem{} Schramm D. N., Turner M. S., 1998, 
Rev. Mod. Phys. 70, 303

\bibitem{} Sembach K. R., Savage B. D.,
Lu L., Murphy E. M., 1999, ApJ 515, 108

\bibitem{} Silk J., 1996, in: Cosmologie
et structure \`a grande \'echelle (Cosmology and large scale structure),
R. Schaeffer, J. Silk, M.Spiro, J. Zinn-Justin, eds., Elsevier,
Amsterdam, p.\ 75

\bibitem{} Tenorio-Tagle G., Franco J.,
Bodenheimer P., R\'o\`zyczka M., 1987, A\&A 179, 219

\bibitem{} Tenorio-Tagle G., Franco J.,
Bodenheimer P., R\'o\`zyczka M., 1988, A\&A 193, 372 (erratum)

\bibitem{} Tufte S. L., Reynolds R. J.,
Haffner L. M., 1998, ApJ 504, 773

\bibitem{} Vaughan A. H., Preston G. W., 1980,
PASP 92, 385

\bibitem{} Verschuur G. L., 1969, ApJ 156, 771

\bibitem{} Verschuur G. L., 1975, ARA\&A 13, 257

\bibitem{} Wakker B. P., 1991, A\&A 250, 499

\bibitem{} Wakker B. P., van Woerden H., 1997,
ARA\&A 35, 217

\bibitem{} Wakker B. P., Howk J. C., Savage B. D., et al.,
1999a, in: Stromlo Workshop on High-Velocity Clouds, 
eds. Gibson, B. K., Putman, M. E., ASP Conference Series Vol. 166, p. 26

\bibitem{} Wakker B. P., van Woerden H., 
Gibson B. K., 1999b, in: Stromlo Workshop on High-Velocity Clouds, 
eds. Gibson, B. K., Putman, M. E., ASP Conference Series Vol. 166, p. 311

\end{thebibliography}
\end{document}